\documentclass[english,times,pra,twocolumn,superscriptaddress]{revtex4-2}
\usepackage[T1]{fontenc}
\usepackage[latin9]{inputenc}
\usepackage{amsmath}
\usepackage{stmaryrd}
\usepackage{graphicx}

\makeatletter
\usepackage{times}

\makeatother

\usepackage{babel}
\begin{document}
\title{Phase-locked amplification enhanced by spin squeezing }
\author{Yan Zhang}
\affiliation{Institute of Applied Physics and Materials Engineering, University
of Macau, Macau}
\author{Jing Zhang}
\affiliation{School of Automation Science and Engineering, Xi\textquoteright an
Jiaotong University, Xi\textquoteright an, 710049, China.}
\affiliation{MOE Key Lab for Intelligent Networks and Network Security, Xi\textquoteright an
Jiaotong University, Xi\textquoteright an 710049, China.}
\author{Hou Ian}
\email{houian@um.edu.mo}

\affiliation{Institute of Applied Physics and Materials Engineering, University
of Macau, Macau}
\begin{abstract}
Quantum lock-in amplification raises the detection sensitivity of
magnetic fields to unprecedented levels by phase-locked pumping the
Zeeman levels of a single trapped atom. However, random spin precessions
limits the useful detection range of arming times for locking high-contrast
signals. To extend this range imposed by the uncertainty limit, quadrature
spin squeezing can be introduced, on top of the phase-locking mechanism.
We propose a detection scheme using an atomic ensemble whose collected
spin is pumped by two lasers for simultaneous squeezing and phase
locking. We derive the optimal $\pi/2$-pulse and $\pi$-pulse schemes
that accomplishes this concurrent action and prove that the resulting
phase sensitivity is enhanced while the usable detection window for
phase locking is widened.
\end{abstract}
\maketitle

\section{Introduction}

Quantum metrology is a discipline that studies how to use the principles
of quantum mechanics to improve measurement accuracy and sensitivity.
It involves theoretical frameworks and experimental methods for designing
and analyzing quantum measurement processes. Among them, quantum sensors
are technologies used in the field of quantum metrology to achieve
high-precision measurements. They can detect and measure weak physical
and chemical signals and have broad application prospects. The excellent
sensitivity offered by quantum sensors has been a major driver of
developments in the fields~\citep{Moreva2020,Yu2020,Xiao2022}. As
a quantum sensor, it is required that the quantum sensor has a strong
response to the useful signal on the one hand, and minimizes the influence
of unwanted noise on the other hand. Quantum metrology aims to improve
the accuracy of measurements by reducing their fundamental statistical
uncertainty given by quantum fluctuations. In atomic interference
measurements, the ultimate limit of precision is subject to quantum
projection noise in measurements of a collection of uncorrelated particles~\citep{Yurke1986,YurkeB1986}.
This noise establishes the standard quantum limit (SQL) of measurement
precision in experiments. SQL is a basic concept in quantum measurement,
which describes the lowest noise level that can be achieved in the
measurement of certain physical quantities under the framework of
quantum mechanics~\citep{Itano1993}. This limit is often associated
with the Heisenberg uncertainty principle. In quantum optics and quantum
metrology, the SQL usually refers to the minimum quantum noise level
that can be achieved without using squeezed states of light. When
it comes to the phase measurement, the SQL of the phase sensitivity
is given by $\delta\phi=1/\sqrt{N}$ in $N$-atom ensemble~\citep{Kitagawa1993}.
Quantum lock-in amplifier can greatly suppress noise and improve the
detection signal to noise ratio. Moreover, it has a very high detection
sensitivity and a relatively simple signal processing, which is an
effective method for weak signal detection~\citep{Kotler2011,Zhuang2021,Jose2023}.
For a single-ion quantum lock-in amplifier, a single trapped ion is
used as a two-level quantum probe, and a train of $N$ $\pi$-pulses
is applied to improve the lock-in measurement sensitivity. Related
experiments have already approved that the quantum lock-in amplifier
can effectively increase the sensitivity of any quantum sensor~\citep{Kotler2011,Kotler2013}.
Inspired by the above, we explored quantum phase locked technology
in multi-particle systems to see if we can further improve phase sensitivity.

Quantum sensors use the properties of quantum states to improve the
accuracy and sensitivity of measurements, and quantum spin squeezing
is a specific quantum state that can reduce the uncertainty in quantum
systems, thereby surpassing the SQL and playing an important role
in quantum precision measurement. In recent years, atomic spin squeezing
has made experimental breakthroughs in the fields of optical atomic
clocks and interferometers, which has well verified theoretical expectations~\citep{Pedrozo-Pe=0000F1afiel2020,Greve2022}.
It also has attracted broad attention for its applications of improving
the precision of measurements~\citep{Braverman2019,Doering2010,Bao2020}
and generating many-body entanglement~\citep{Perlin2020,Amico2008,Comparin2022}.
Using quantum spin-based systems, the uncertainty of the atomic ensemble
can be redistributed, so that the uncertainty of observables related
to measurement is reduced, while the uncertainty of parameters unrelated
to measurement is increased, thereby breaking through the SQL and
ultimately approaching the Heisenberg limit of quantum mechanics.
Hence, the concept of spin squeezed states (SSS) consisting of many-body
entangled states were proposed to overcome the SQL~\citep{Kitagawa1993,Pezze2018}.
Here, we use a one-axis twisting mechanism to generate spin squeezing
via atom--photon interactions in $N$-atom ensemble~\citep{Huang2021,Sanders1989,Opatrn=0000FD2018}.
Since spin squeezing can reduce phase sensitivity without violating
the uncertainty relationship, it can increase measurement sensitivity
of a lock-in amplifier. Considering that SSS are generated in a bunch
of atomic ensembles, we load quantum phase locked signals into a multi-particle
system instead of a single two-level system.

In quantum optics, nonlinear Hamiltonians can SSS, where the nonlinear
Kerr effect is used to prepare squeezed light. Quantum SSS usually
involve quantum entanglement, which is a non-classical correlation
between quantum systems. The existence of entanglement makes it impossible
to describe the overall properties of a quantum system by the properties
of a single particle alone, which is the key factor in achieving quantum
spin squeezing. We use photons denoted by Stokes operators $S_{z}$
to induce squeezing in atomic ensembles denoted by collective spin
operators $J_{z}$ through their interaction Hamiltonian. The specific
implementation of the squeezing effect $J^{2}_{z}$ in atomic ensembles
is achieved via a train of $\pi/2$ pulses separated with free evolutions.
In order to combine the squeezed atomic ensemble with quantum phase
locked technology, we expanded the single two-level system in the
original phase locked amplifier into a multi-particle system. The
total Hamiltonia consists of the effective squeezed part, a modulated
signal part and a train of lock-in $\pi$ pulse part. According to
the original definition of minimal detectible phase based on the Ramsey
fringe signal $\delta\phi=\left(\Delta J^{2}_{\perp}\right)^{1/2}/\left\langle J\right\rangle $~\citep{Wineland1994,Andre2002,Ma2011},
we deduced $\left(\Delta J^{2}_{z}\right)^{1/2}$ and $\left\langle J_{x}\right\rangle $
after the Hamiltonian evolution operator acting on the initial coherent
spin state, and concluded that the spin squeezed atomic ensembles
can help further improve the phase sensitivity in the phase locked
amplification.

The main contents of the paper are as follows. In Sec.~.\ref{sec:model},
we describe the system Hamiltonian in detail, including how SSS of
the atomic ensemble are generated.

\section{Spin squeezing for phase locking\label{sec:model}}

\begin{figure}
\includegraphics[clip,width=8.6cm]{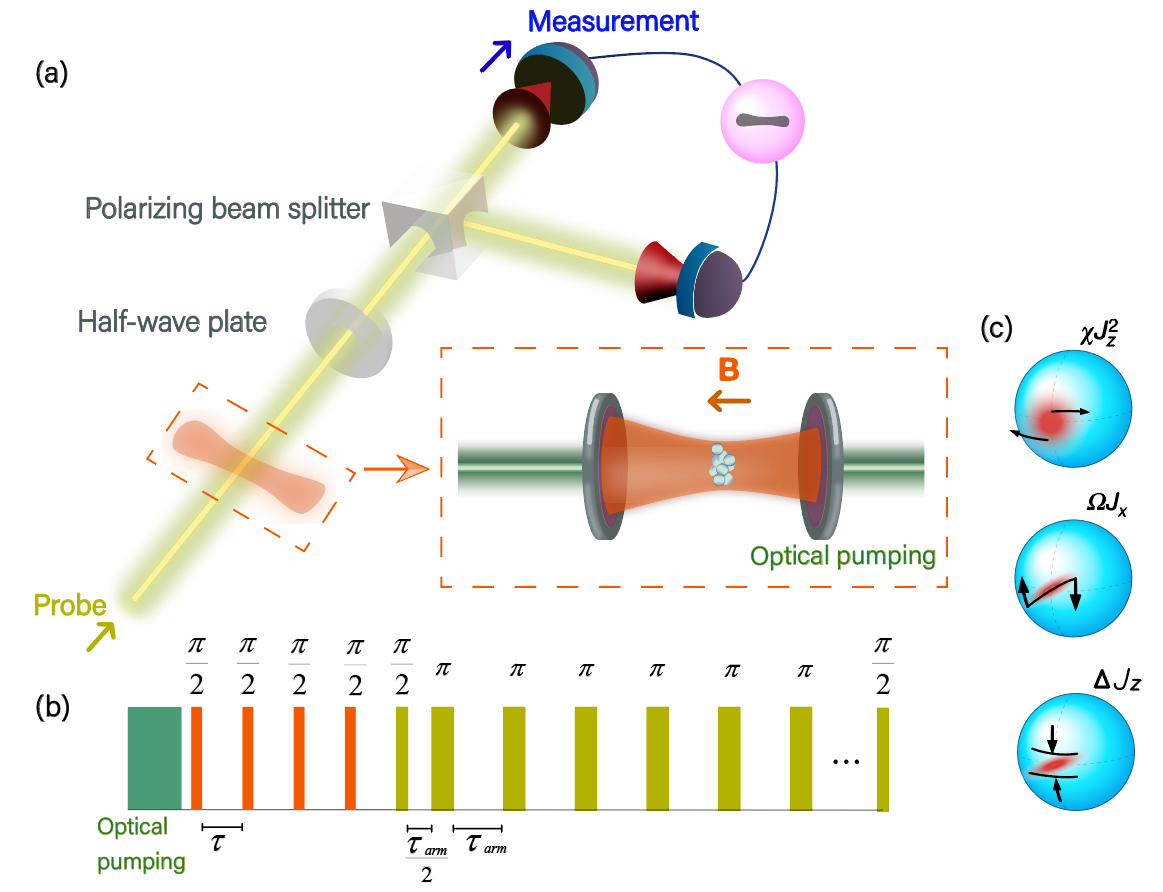}\caption{\label{fig:Setup}Experimental setup. (a)Schematic of the setup. Atoms
are trapped inside an optical cavity. By inserting a half-wave plate
(HWP) in front of the polarizing beam splitter (PBS), the probe beam
is split into $x$- and $y$-polarized beams in the output ports of
PBS and then detected by two photon detectors. (b)Pulse sequence.
The atoms are initially prepared in a CSS by optical pumping propagating
along the $x$ direction. The orange pulse train indicates the squeezing
pulse scheme, and the yellow-green pulse train indicates the quantum
phase locked measurement pulse scheme.}
\end{figure}

Quantum metrology makes use of a specific quantum system that interacts
with and thus changes its state along with the signal to be detected.
Through the readout of the final state of that system and the data
post-processing process, the signal in question can be detected with
ultra high sensitivity limited by the principles of quantum physics.
Among this type of techniques, quantum lock-in amplification probes
the signal from a noise floor, using a microwave driven two-level
atom. The periodically beaten atom spin locks and trails the phase
of the signal while being measured by a laser beam, resulting in a
readout sensitivity limited only by the quadrature uncertainty and
the locking quality. In quantum SSS, the uncertainty of a collective
spin component can be further squeezed below the uncertainty limit
of a single spin by sacrificing the uncertainty of another (unused)
orthogonal spin component. We explore the phase-locked amplification
technique with simultaneous spin squeezing that would lead to further
enhancement in phase sensitivity. Consider a system of $N_{a}$ two-level
atoms interacting with a low-frequency magnetic field, a driving laser
beam and a microwave driving field, as illustrated in Fig.~\ref{fig:Setup}(a). 

The atoms are assumed degenerate originally and the interacting magnetic
field lifts the degeneracy by splitting the fine levels such that
the level spacing is depending on the magnetic field strength. In
other words, the free energy of the atoms is 
\begin{equation}
H_{a}=\frac{1}{2}\sum^{N_{a}}_{j}M(t)\sigma_{j,z}=M(t)J_{z},\label{eq:Ham_atoms}
\end{equation}
presuming equal coupling for each of the atoms. The collective atomic
ensemble consisting of $N_{a}$ half spins is described by the total
angular momentum operators $J_{\mu}=\sum_{j}\sigma_{j,\mu}/2$, where
$\sigma_{j,\mu}$ is the Pauli $\mu$-matrix ($\mu\in\{x,y,z\}$)
for the $j$-th atom. The signal $M(t)$ measured by the detector
consists of the modulated signal $S(t)$ and the noise $N(t)$, i.e.
$M(t)=S(t)+N(t)$.

The microwave driving intends to locked the magnetic field signal
by periodically driving the atoms semi-classically with specifically
spaced pulses, giving the time-dependent driving Hamiltonian 
\begin{equation}
H_{d}=\frac{1}{2}\sum^{N_{a}}_{j}\sum^{N}_{m}\Omega_{0}\delta(t-m\tau_{\mathrm{arm}})\sigma_{j,x}=\Omega(t)J_{x}\label{eq:Ham_drive}
\end{equation}
where $\Omega(t)$ indicates a train of $N$ equally $\tau_{\mathrm{arm}}$-spaced
$\pi$-pulses of driving strength $\Omega_{0}$ (illustrated by the
yellow-green pulses in Fig.~\ref{fig:Setup}(b)). The laser beams,
meanwhile, squeezes all the atoms as a coherent atomic ensemble to
further lower the noise level of the atomic vibrations such that the
atomic ensemble can act as an ultra low noise amplifier. The laser-atom
coupling is described by the Faraday Hamiltonian 
\begin{equation}
H_{\mathrm{int}}=gJ_{z}S_{z}\label{eq:Faraday Ham}
\end{equation}
where the photon spins are represented by the Stokes operator $S_{z}=(a^{\dagger}_{x}a_{y}-a^{\dagger}_{y}a_{x})/2i$~\citep{Bao2020},
with $\{a^{\dagger}_{\mu},a_{\mu}\}$ being the creation and annihilation
operator pair in the $\mu=x$ or $y$ polarization.

Initially, the interaction Hamiltonian is aligned along the $z$-axis,
with the initial state being a coherent spin state (CSS) pointing
along the $x$-axis. The photon total spin $S$ act as an intermediary
to induce a torque on atomic spin $J$ to transfer the angular momentum.
To generate the desired squeezing, a train of four $\pi/2$ pulses
separated with free evolutions is applied on the photons, as illustrated
by the orange pulse train in Fig.~\ref{fig:Setup}(b). The evolution
operator describes as 
\begin{equation}
U_{4}(\tau)=\left[R_{S}\left(\frac{\pi}{2}\right)U(\tau)\right]^{4}\label{eq:U_4}
\end{equation}
where $R_{S}(\pi/2)=e^{-i\frac{\pi}{2}S_{x}}$ is the $\pi/2$ rotation
pulse along $x$- axis, $S_{x}=\frac{1}{2}(a^{\dagger}_{x}a_{x}-a^{\dagger}_{y}a_{y})$
is the Stokes operator. $U(\tau)=\exp\{-igJ_{z}S_{z}\tau\}$ is the
free evolution operator according to the Hamiltonian in Eq.~(\ref{eq:Faraday Ham}).
With the Baker-Campbell-Hausdorff formula, we rewrote Eq.~(\ref{eq:U_4})
after the operator expansion as 

\begin{equation}
U^{\prime}_{4}(\tau)=\exp\left\{ -i(g\tau)^{2}S_{x}J^{2}_{z}+\frac{i}{2}(g\tau)^{3}(S_{y}-S_{z})J^{3}_{z}+\frac{2i}{3!}(g\tau)^{4}S_{x}J^{4}_{z}+\dots\right\} .\label{eq:U_4'}
\end{equation}
Since our system is under the week coupling regime, the high-order
terms of $(g\tau)^{3}$,$(g\tau)^{4}$, etc. can be ignored. The second-order
term contributes the nonlinear interaction $J^{2}_{z}$, which leads
to the one-axis twisting. Then, we simplify the effective Hamiltonian
as
\begin{equation}
H^{\prime}_{\mathrm{eff}}=\frac{1}{4}g{}^{2}\tau S_{x}J^{2}_{z}.\label{eq:squeezed H}
\end{equation}
By the nonlinear interaction in Eq.~(\ref{eq:squeezed H}), the atomic
ensemble is already been squeezed. The angular momentum operator $J^{2}_{z}$
donates the spin squeezing generated in the atomic ensemble, which
twists the quantum fluctuations. Since the initial state is polarized
along the $x$-axis, it is the eigenstate of $S_{x}$ with maximum
eigenvalue is half of the number of photons $N_{s}/2$. Thus, the
effective Hamiltonian is rewritten as
\begin{equation}
H_{\mathrm{eff}}=\chi J^{2}_{z}\label{eq:eff H}
\end{equation}
where $\chi=N_{s}g^{2}\tau/8$ is characterised as the effective interaction
strength. $\chi$ determines how uncertainties are deformed by twisting.
Finally, the system Hamiltonian formed is 
\begin{equation}
H=\chi J^{2}_{z}+M(t)J_{z}+\Omega(t)J_{x}.\label{eq:sys H}
\end{equation}

\section{Fringe contrast\label{sec:contrast}}

We assume that the atoms are initially polarized along the $x$-axis
and prepared in a coherent spin state (CSS) by optical pumping, as
shown by the green pulse in Fig.~\ref{fig:Setup}(b). The pure state
of $N_{a}$ spin-$\frac{1}{2}$ system prepared in the CSS is described
as

\begin{equation}
\left|\theta_{k},\phi_{k}\right\rangle _{k}=\cos\frac{\theta_{k}}{2}\left|\uparrow\right\rangle _{k}+e^{i\phi_{k}}\sin\frac{\theta_{k}}{2}\left|\downarrow\right\rangle _{k}\label{eq:css}
\end{equation}
where $\left|\uparrow\right\rangle _{k}(\left|\downarrow\right\rangle _{k})$
is the eigenstate of $J_{z}$ with the eigenvalue $\frac{1}{2}(-\frac{1}{2})$
and $\theta_{k}(\phi_{k})$ describes elevation (azimuth) in the $k$th
spin-$\frac{1}{2}$ system. 

Since the atoms $J$ are along the $z$-axis direction, to ensure
that effective changes are generated, we assume the initial the mean-spin
direction falls on the $x$-axis with all $\theta_{k}=\frac{\pi}{2},\phi_{k}=0$.
The initial CSS is tensor product
\begin{align}
\left|\theta_{k}=\frac{\pi}{2},\phi_{k}=0\right\rangle  & =\left(\frac{\sqrt{2}}{2}\left|\uparrow\right\rangle _{1}+\frac{\sqrt{2}}{2}\left|\downarrow\right\rangle _{1}\right)\varotimes\nonumber \\
 & \cdots\varotimes\left(\frac{\sqrt{2}}{2}\left|\uparrow\right\rangle _{N_{a}}+\frac{\sqrt{2}}{2}\left|\downarrow\right\rangle _{N_{a}}\right).\label{eq:ini css}
\end{align}
In the quantum lock-in amplifier, the measurement sensitivity is characterized
specifically as the phase sensitivity and is directly related to the
minimal detectible phase given by $\delta\phi=\left(\varDelta J^{2}_{z}\right)^{1/2}/\left\langle J_{x}\right\rangle $\citep{Wineland1994}.
Since $J=\frac{1}{2}\sum^{N_{a}}_{i=1}\sigma_{i}$ , consequently
we need to know the expectation of $\sigma_{i,k}$. Here, $i$ is
the index of the $i$-th of the atoms and $k$ is the index of the
$\sigma_{x},\sigma_{y},\sigma_{z}$ Pauli matrices. The evolution
operator according to the system Hamiltonian in Eq.~(\ref{eq:sys H})
can be written as 
\begin{equation}
U=\exp-i\left\{ \alpha J^{2}_{z}+\beta J_{z}+\gamma J_{x}\right\} \label{eq: evolution operator}
\end{equation}
where $\alpha=\chi t,\beta=\int^{t}_{0}M(t')dt',\gamma=\int^{t}_{0}\varOmega(t')dt'$
are the average phase accumulated over the period of $t$. Therefore,
the evolution of $\sigma_{i,x}$ can be obtained as $U^{\dagger}\sigma_{i,x}U$,
then the expectation $\left\langle \sigma_{i,x}\right\rangle $ is
\begin{align}
\left\langle \sigma_{i,x}\right\rangle  & =\cos\left(\alpha J^{\prime}_{z}\right)\left[\sigma_{i,x}\cos\left(\beta\right)-\sigma_{i,y}\sin\left(\beta\right)\right]\nonumber \\
 & -\sin\left(\alpha J^{\prime}_{z}\right)\left[\sigma_{i,y}\cos\left(\beta\right)+\sigma_{i,x}\sin\left(\beta\right)\right]\label{eq:expect sigma_ix}
\end{align}
in which $J^{\prime}_{z}$ is the operator defined as 
\begin{equation}
J^{\prime}_{z}=\sum^{N_{a}}_{m\neq i}\sigma_{m,z}.\label{eq:J_z'}
\end{equation}
$m$ is the index of the $m$-th of the atoms except the $i$-th atom.
Using
\begin{align}
\left\langle \frac{\pi}{2},0\right|\sigma_{i,x}\left|\frac{\pi}{2},0\right\rangle  & =1,\label{eq:sig_ix}\\
\left\langle \frac{\pi}{2},0\right|\sigma_{i,y}\left|\frac{\pi}{2},0\right\rangle  & =\left\langle \frac{\pi}{2},0\right|\sigma_{i,z}\left|\frac{\pi}{2},0\right\rangle =0\label{eq:sig_iy}\\
\left\langle \frac{\pi}{2},0\right|\cos\left(\alpha J^{\prime}_{z}\right)\left|\frac{\pi}{2},0\right\rangle  & =\cos^{N_{a}-1}\alpha,\label{eq:cosJz'}\\
\left\langle \frac{\pi}{2},0\right|\sin\left(\alpha J^{\prime}_{z}\right)\left|\frac{\pi}{2},0\right\rangle  & =\sin^{N_{a}-1}\alpha,\label{eq:sinJz'}
\end{align}
in Eq.~(\ref{eq:expect sigma_ix}), we get the expectation $\left\langle J_{x}\right\rangle $
,
\begin{equation}
\left\langle J_{x}\right\rangle =\frac{N_{a}}{2}\cos^{N_{a}-1}\alpha\cos\beta-\frac{N_{a}}{2}\sin^{N_{a}-1}\alpha\sin\beta\label{eq:expect Jx}
\end{equation}

To deduce $\left(\Delta J^{2}_{z}\right)^{1/2}$, we need to know
the expectation $\left\langle \sigma_{i,z}\right\rangle $ to obtain
$\left\langle J_{z}\right\rangle $. With the system evolution operator
in Eq.(\ref{eq: evolution operator}), we have the expectation $\left\langle \sigma_{i,z}\right\rangle $
\begin{align}
\left\langle \sigma_{i,z}\right\rangle  & =\cos^{N_{a}-1}\alpha\sin\beta\sin\gamma+\sin^{N_{a}-1}\alpha\cos\beta\sin\gamma\label{eq:expect sigma_iz}
\end{align}
With Eq.~(\ref{eq:sig_ix}), (\ref{eq:sig_iy}), (\ref{eq:cosJz'})
and (\ref{eq:sinJz'}), we solve the expectation $\left\langle J_{z}\right\rangle $
from Eq.~(\ref{eq:expect sigma_iz})
\begin{align}
\left\langle J_{z}\right\rangle  & =\frac{N_{a}}{2}\cos^{N_{a}-1}\alpha\sin\beta\sin\gamma+\frac{N_{a}}{2}\sin^{N_{a}-1}\alpha\cos\beta\sin\gamma.\label{eq:expect Jz}
\end{align}
As for $\left\langle J^{2}_{z}\right\rangle $, following the similar
solution procrdure with $\left\langle J_{x}\right\rangle $ and $\left\langle J_{y}\right\rangle $,
we have
\begin{align}
\left\langle J^{2}_{z}\right\rangle  & =\frac{N_{a}}{4}.\label{eq:expect Jz^2}
\end{align}
Finally, by Eq.~(\ref{eq:expect Jx}), (\ref{eq:expect Jz}), and
(\ref{eq:expect Jz^2}), we obtain the minimal detectible phase
\begin{equation}
\delta\phi=\frac{\sqrt{\frac{1}{N_{a}}-\left(\cos^{N_{a}-1}\alpha\sin\beta\sin\gamma+\sin^{N_{a}-1}\alpha\cos\beta\sin\gamma\right)^{2}}}{\cos\beta\cos^{N_{a}-1}\alpha-\sin\beta\sin^{N_{a}-1}\alpha}.\label{eq:phase diff}
\end{equation}

\begin{figure}
\includegraphics[bb=20bp 10bp 390bp 320bp,clip,width=8.6cm]{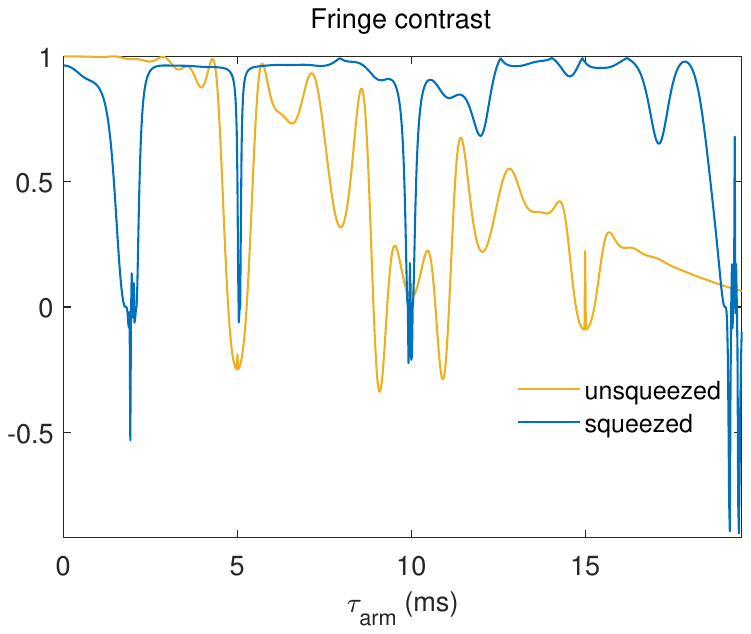}\caption{\label{fig:Contrast compare}Fringe contrast with spin squeezed and
without squeezed. Photon number and atom number are $N_{s}=N_{a}=50$
and squeezing pulse separation is $\tau=1\times10^{-4}g^{-1}$, the
effective interaction strength is set to $\chi=6.25\times10^{-4}g$. }
\end{figure}

\begin{figure}
\includegraphics[bb=20bp 8bp 400bp 315bp,clip,width=8.6cm]{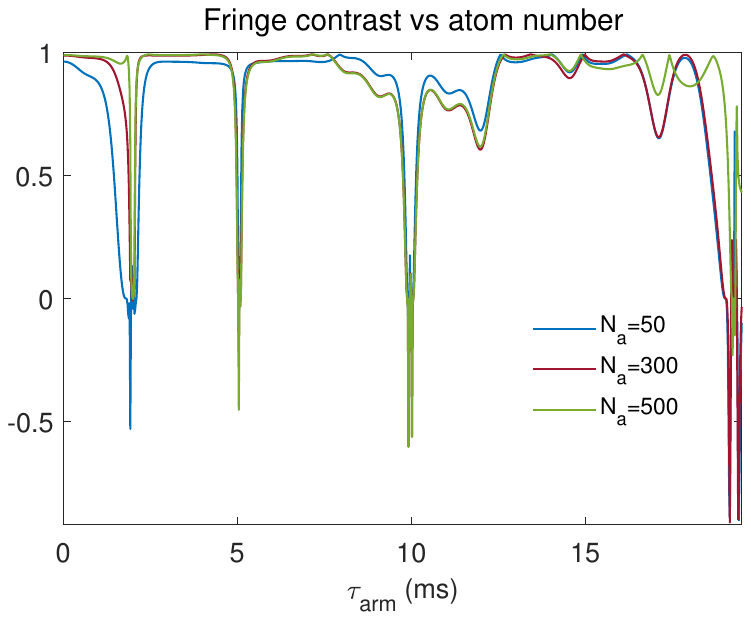}\caption{\label{fig:Contrast n_atom}Comparison of fringe contrast versus the
number of atoms. }
\end{figure}

To get the measurement sensitivity of the phase locked amplification,
we first measure the fringe contrast in the absence of any modulated
signal. Assume that there are $q$ discrete magnetic field noise components,
and the time domain noise can be obtained by inverse Fourier transform
\begin{equation}
N(t)=\Sigma^{q}_{k=1}|N_{k}|\cos(\theta_{k}+2\pi f_{k}t).\label{eq:Noise_t}
\end{equation}
Then the accumulated phase $\beta$ is 
\begin{equation}
\beta=\Sigma^{q}_{k=1}\frac{|N_{k}|}{2\pi f_{k}}\sin(\theta_{k}+2\pi f_{k}t)
\end{equation}
where $\theta_{k}$ is uniformly distributed in $[0,2\pi]$. We assume
that averaging repeated quantum projection measurements is equivalent
to treating $\theta_{k}$ as an independent random variable, so the
fringe contrast is 
\begin{equation}
A=\int\cos(\delta\phi)d^{q}\theta_{k}.\label{eq:contrast}
\end{equation}
In order to verify the effectiveness of the algorithm, we assume that
there are three magnetic noise spectral components, i.e., 50Hz, 100Hz
and $f_{\mathrm{slow}}$, where $f_{\mathrm{slow}}=2.1\mathrm{Hz}$
represents a slowly varying field. The respective noise amplitudes
are $B_{50\mathrm{Hz}}=540\mathrm{pT},\:B_{100\mathrm{Hz}}=390\mathrm{pT}$
and $g_{e}\mu_{B}B_{\mathrm{slow}}f_{\mathrm{slow}}/h=40\mathrm{Hz^{2}}$\citep{Kotler2011}.
Fig.~\ref{fig:Contrast compare} shows the fringe contrast $A$ versus
the pulse interval (arming time) $\tau_{\mathrm{arm}}$ for $N=7$
pulses per phase locked sequence. As shown, we can see that there
are dips at $\tau_{\mathrm{arm}}=5\mathrm{ms}$ and $10\mathrm{ms}$
corresponding to the assumed $100\mathrm{Hz}$ and $50\mathrm{Hz}$
magnetic noise components. Ideally, according to the Ramsey spectroscopy
measurement, the closer the value of the fringe contrast $A$ is to
1, the closer the measurement effect is to the theoretical value.
Comparing the two curves in Fig.~\ref{fig:Contrast compare}, the
fringe contrast with spin squeezing is far better than that of the
unsqueezed. Obviously, the smooth range of the squeezed curve close
to 1 is significantly wider than that of the unsqueezed curve. Next,
in order to better highlight the impact of squeezing on the measurement
of the fringe contrast, we draw a comparison Fig.~\ref{fig:Contrast n_atom}
of the fringe contrast as the number of atoms increases. The range
of the green curve close to 1 is wider than that of the other two
curves, i.e. from $10$ to $19.24$~ms. We call this range the experimental
measurement range. Although the red curve and the blue curve have
the same measurement range, i.e., from $10$ to $19.16$~ms. the
maximum value that the red curve can reach is closer to 1 than the
blue one.

\section{Improvement of sensitivity\label{sec:sensitivity}}

\begin{figure}
\includegraphics[bb=30bp 0bp 400bp 310bp,clip,width=8.6cm]{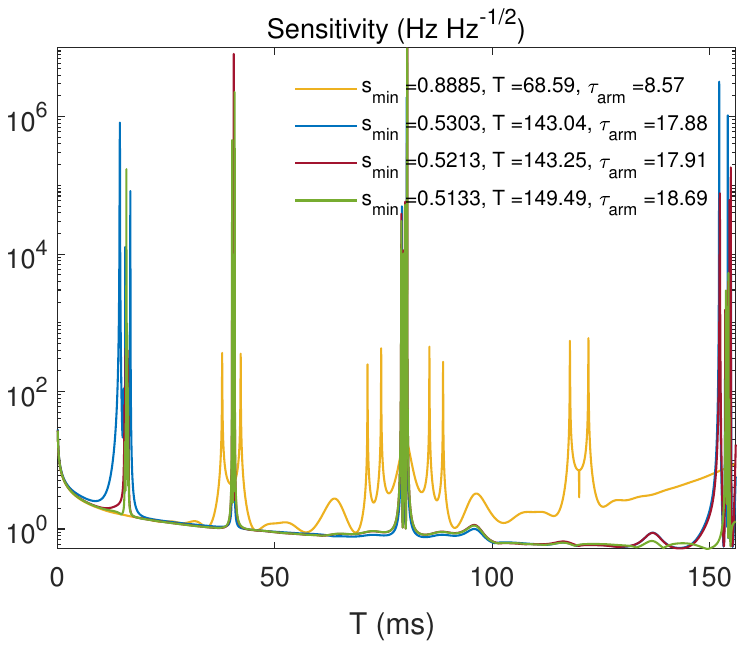}\caption{\label{fig:sensitivity}Sensitivity (solid blue line) versus the phase
locked sequence duration $T$.}
\end{figure}

The quantum phase locked amplification works similarly to the classical
locked amplification, but takes advantage of the properties of quantum
systems. Taking a single Sr\textsuperscript{+} ion as an example,
the quantum state is initialized and measured through optical pumping
and selective fluorescence technology, and the ion probe is modulated
using a train of $\pi$ pulse sequence to achieve high-precision measurement
of weak signals. On this basis, we propose to use the spin squeezed
atomic ensemble as a probe to see whether the measurement sensitivity
can be further improved. Fig.~\ref{fig:sensitivity} shows the sensitivity
comparisons of unsqueezed, 50-atom squeezed, 300-atom squeezed, and
500-atom squeezed. It can be seen that as the number of squeezed atoms
increases, the measurement sensitivity will increase, and the experimental
time of one phase locked sequence duration will increase accordingly.
A best sensitivity of 0.5133 $\mathrm{Hz\,Hz^{-1/2}}$ is obtained
of the green curve at T=149.49 ms.

\section{Conclusions\label{sec:conclusions}}

We use quantum spin squeezing technology to further improve the phase
sensitivity of phase locked amplification. Quantum spin squeezing
can redistribute the uncertainty of an atomic collection, reducing
the uncertainty of observables related to measurement while increasing
the uncertainty of parameters unrelated to measurement, thereby further
improving the phase locked amplification. We expand the single two-level
system in the original phase locked amplification into a multi-particle
system to combine atomic ensemble SSS. The comparison proves the effectiveness
of this strategy in improving phase sensitivity.
\begin{acknowledgments}
H. I. acknowledges the support by FDCT of Macau under grants 0015/2021/AGJ
and 0134/2024/AFJ and by the Guangdong Provincial Quantum Science
Strategic Initiative (Grants No. GDZX2203001, GDZX2403001).
\end{acknowledgments}

\end{document}